\documentclass[11pt,twoside]{article}

\usepackage{asp2006}
\usepackage{epsf}
\usepackage{psfig}
\usepackage{lscape}
\usepackage{epsfig,subfigure,amsmath,amssymb}

\begin{document}

\title{12 GHz Radio-Holographic surface measurement of the RRI 10.4 m telescope}
\author{Ramesh Balasubramanyam, Suresh Venkatesh and Sharath B. Raju}
\affil{Raman Research Institute, Bangalore 560 080, INDIA\\
%\author{Ramesh Balasubramanyam, Suresh Venkatesh, Sharath B. Raju}
%Raman Research Institute, Bangalore 560 080, INDIA
%\affil{Raman Research Institute, Bangalore 560 080, INDIA\\
{\tt Email: \{ramesh, sureshv, sharathbr\}@rri.res.in} }

\begin{abstract}
A modern Q-band low noise amplifier (LNA) front-end is being fitted to the 10.4~m
millimeter-wave telescope at the Raman Research Institute (RRI) to support
observations in the 40-50~GHz frequency range. 
To assess the suitability
of the surface for this purpose, we measured the
deviations of the primary surface from an ideal
paraboloid using radio holography. We used the 11.6996 GHz beacon signal from
the GSAT3 satellite, a 1.2~m reference antenna,
commercial K$_{u}$-band Low Noise Block Convereters (LNBC) as the receiver
front-ends and a Stanford Research Systems (SRS)
lock-in amplifier as the backend. The LNBCs
had independent free-running first local oscillators (LO). Yet,
we recovered the correlation by using a radiatively
injected common tone that served as the second local oscillator.
With this setup, we mapped the surface
deviations on a $64 \times 64$ grid and measured an rms
surface deviation of $\sim 350~\mu$m with a measurement
accuracy of $\sim 50~\mu$m.
\end{abstract}

%%%%%%%%%%%%%%%%%%%%%%%%%%%%%%%%%%%%%%%%%%%%%%%%%%%%%%%%%
\section{Introduction}
%\section{Need for surface measurements}
%\label{s:introduction}

RRI has a mm-wave Leighton telescope, of 10.4~m diameter with 81 hexagonal 
panels~\citep{ref:tks}. It is being rejuvenated to undertake Q-band observations
at 43~GHz. This requires the surface rms error to be below
$\sim 370~\mu$m in order to have an aperture efficieny better than 50\%, as
given by Ruze's relation:~$\eta_{ap}=\eta_0 ~exp{(\frac{4\pi \delta}{\lambda})^2}$~\citep{ref:ruze66}.
Radio holographic surface measurement was carried out in Aug-Sep 2007 to 
measure the surface rms error, and to identify panels that may need correction.
In this poster paper, we report the details of this experiment, analyse the data,
discuss the results and present our conclusions.

%Figure: Receiver
\begin{figure}[!t]
\centering
\epsfig{figure=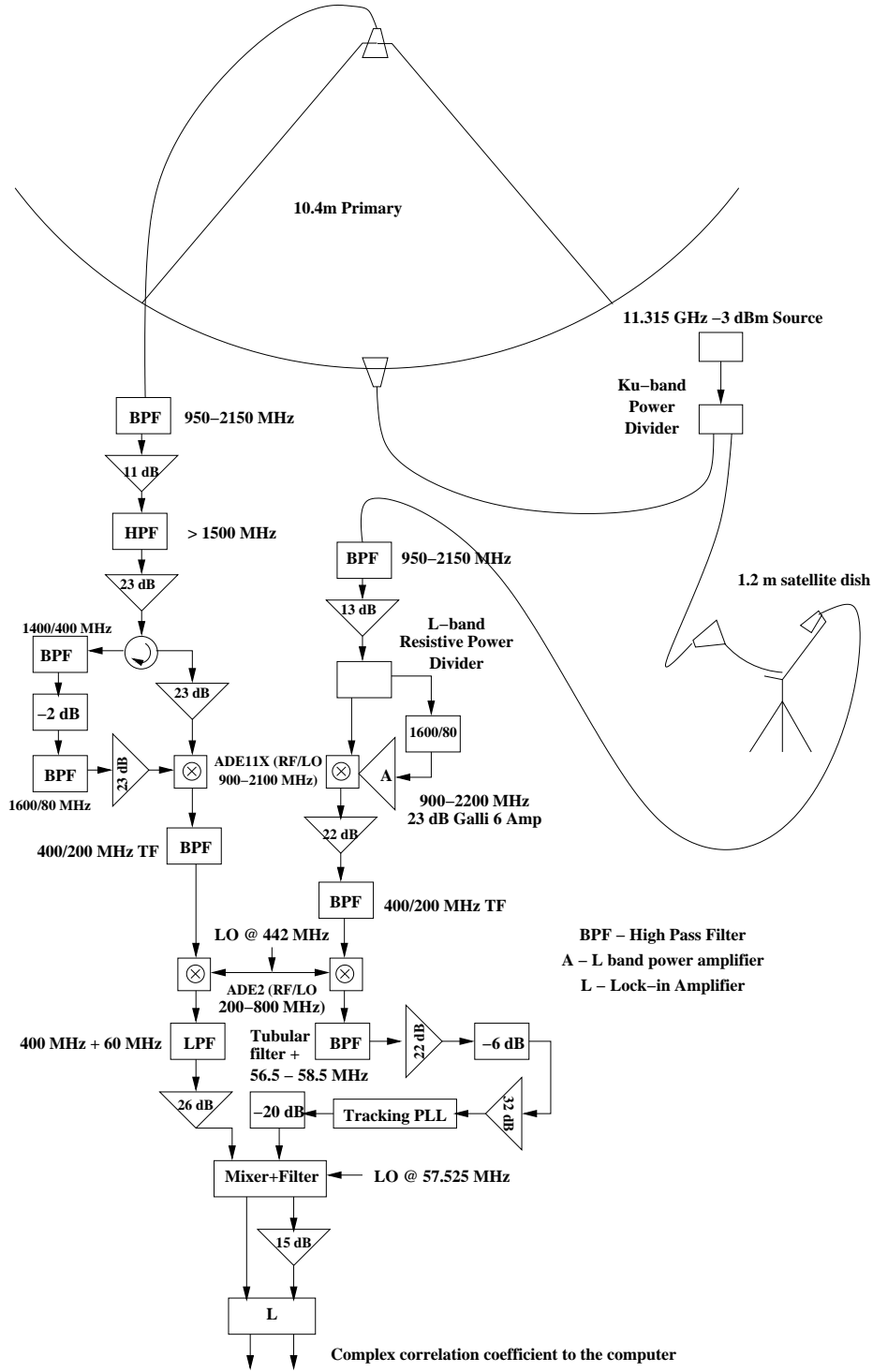, width=0.93\linewidth}
\caption{\small Dual channel holography receiver layout\normalsize }
\label{f:receiver}
\end{figure}
%%%%%%%%%%%%%%%%%%%%%%%%%%%%%%%%%%%%%%%%%%%%%%%%%%%%%%%%%
\section{Experimental setup}

The key idea in Radio Holography is to measure the complex beam pattern. Its
fourier transform yields the aperture plane field (APF) distribution. 
Surface deviations can be calculated from the phases of the APF~\citep{ref:bracewell}.
The signal-to-noise ratio (SNR) required to measure the surface deviation to an accuracy of
 $\delta~\mu$m over $N \times N$ pixels is proportional to $N/\delta$
\citep{ref:scott}. For K$_{u}$-band ($\lambda$~=~25 mm),
the SNR should be $\geq$~1600 to achieve a measurement accuracy of atleast
$80~\mu$m.

Both the test (10.4~m) and the reference (1.2~m)
antennas were fitted with satellite TV low-noise block converters (LNBCs) as
front-ends (see Fig.~\ref{f:receiver}).
A 11.315~GHz reference tone was radiatively
fed to both the antennas for use as second LOs
to remove the effects of the free-running first
LOs in the LNBCs.
A tracking phase locked loop (PLL) improves the SNR by
locking to the signal from the reference
antenna down-converted and filtered to 57.5~MHz.
On both the chains, signals are down-converted to below 100~kHz and fed to the
test (10.4~m) and reference (1.2~m) inputs of the SRS lock-in amplifier (analog correlator).
The correlation amplitudes and phases were digitized with a 12 bit
analog-to-digital converter (ADC) and
recorded in the control computer every 100~ms.

Software was developed to automatically raster-scan the region of interest.
Calibration and scan data from the two position encoders and ADCs were
recorded. In addition, satellite drift had to be compensated at regular intervals.

%%%%%%%%%%%%%%%%%%%%%%%%%%%%%%%%%%%%%%%%%%%%%%%%%%%%%%%%%
\section{Observations}
\begin{figure}[!t]
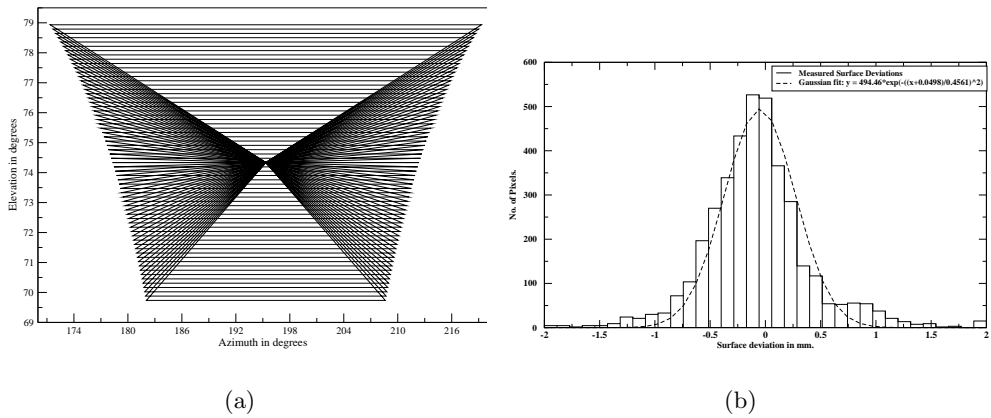

\centering
\subfigure[]{
\epsfig{figure=HoloMap.eps,width=0.48\linewidth}}
\subfigure[]{
\epsfig{figure=hist_surfacedev.eps,width=0.48\linewidth}}
\caption{\small (a) Scheme for raster scan. (b) Histogram of the measured
surface deviations. \normalsize }
\label{f:rasterscan}
\end{figure}

The 11.6996~GHz beacon signal from EDUSAT (Orbital slot: 74$^\circ$°E longitude,
translates to El: 74.40$^\circ$° \& Az: 195.75$^\circ$) was observed. The
correlation co-efficients between the two antennas were recorded on the fly.
The map spans $\pm 4.3^\circ$ in real space in both azimuth (Az) \&
elevation (El) about the satellite position as shown in Fig.~\ref{f:rasterscan}(a).
The critical sampling at 11.7~GHz
is $8.8'$ and the measured beam width is $9.6'$. Since the number of pixels across
the dish is 64 (a map of $64 \times 64$), a spatial resolution of 16~cm was achieved,
which is 1/6th the size of one panel in the dish. Each observation lasted 7-8~hrs
which includes 65 Az scans, 11 satellite pointings and 8 to 9 repeats of central
block of 5 scans. Four independent observations were carried out, all leading to
the same results. All the observations were conducted during night to ensure
system stability.

%%%%%%%%%%%%%%%%%%%%%%%%%%%%%%%%%%%%%%%%%%%%%%%%%%%%%%%%%
%\vspace{-10pt}
\section{Results \& Conclusions}
%\label{s:results}
The effect of free runnning local oscillators in LNBC was solved by radiatively
injecting a tone. The effect of satellite drifting was removed by pointing to the satellite
every 30~minutes. An SNR $>$ 5000 was achieved by making the receiver chain
robust, repeating and co-adding the central block scans \& using satellite
pointing data for amplitude and phase calibration. 

\begin{figure}[!h]
\centering
\subfigure[]{
\epsfig{figure=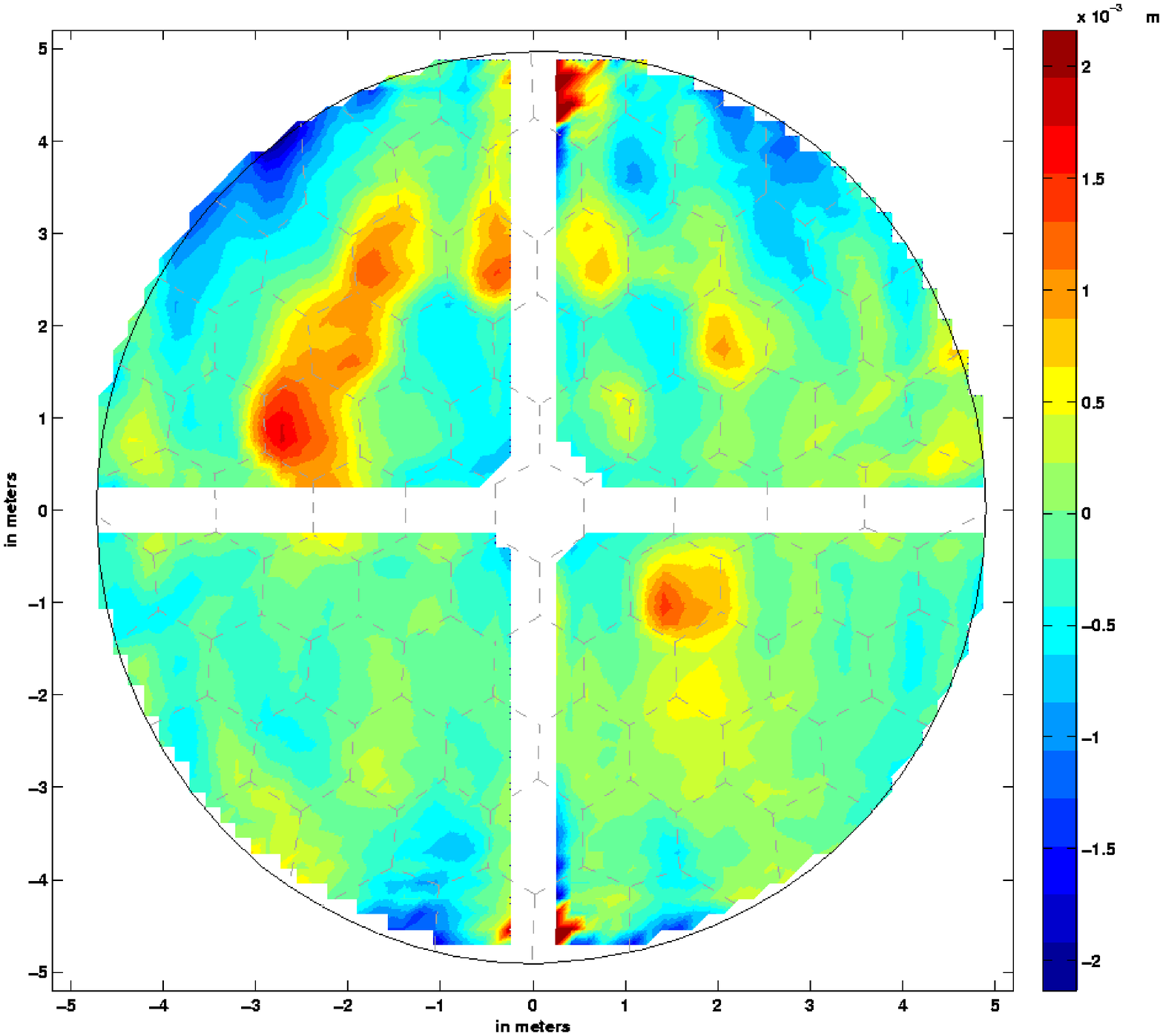,width=0.48\linewidth}}
\subfigure[]{
\epsfig{figure=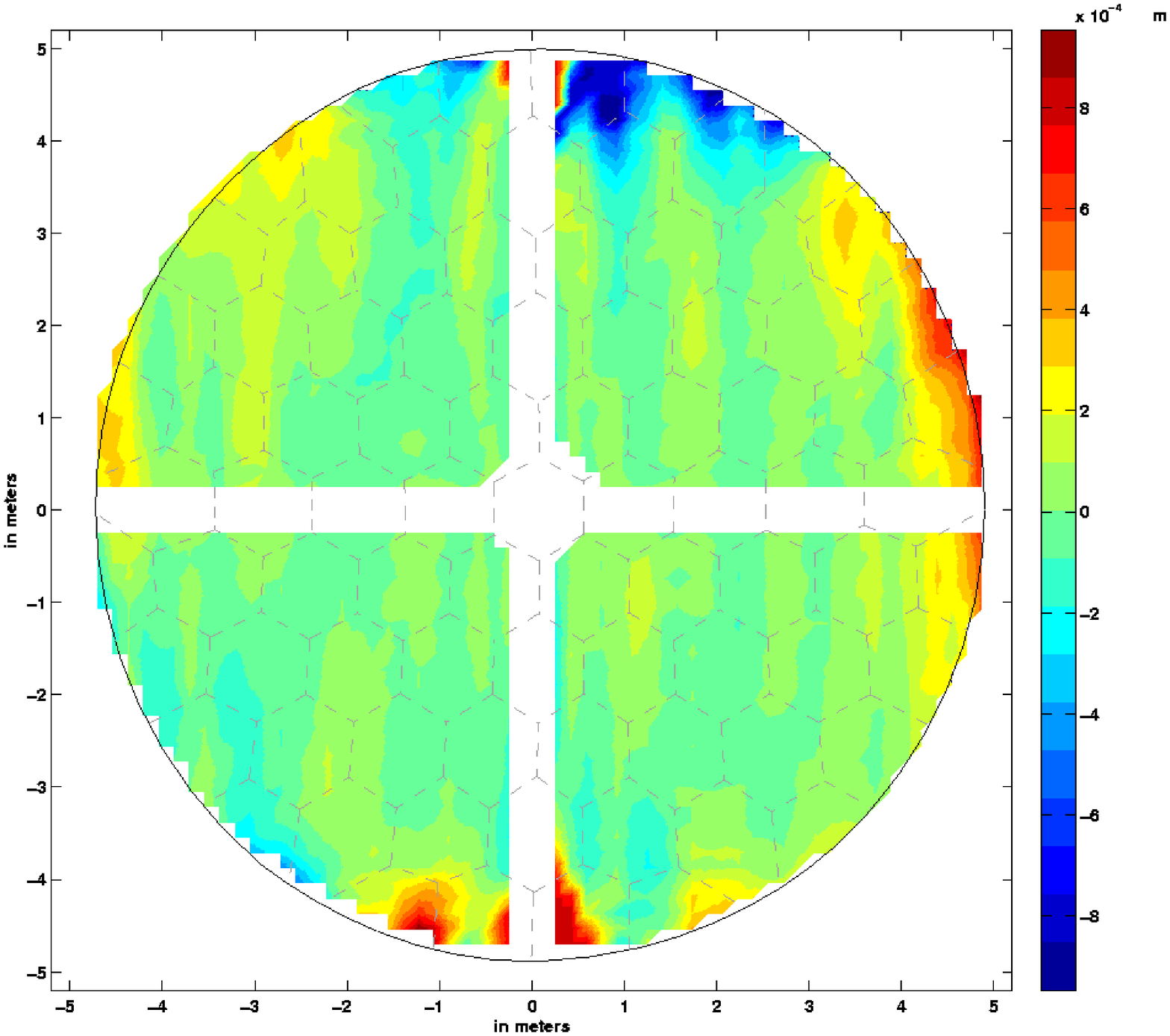,width=0.48\linewidth}}
\caption{\small (a) Measured surface devations. (b) Difference map showing
measurement accuracy\normalsize }
\label{f:map}
\end{figure}

From the measured surface deviations (see Fig.~\ref{f:map}(a)), the
surface rms error is calculated to be $\sim350~\mu$m. It is within $\lambda$/16, implying
that Q Band observations are possible. It can also be seen that some panels
require correction. Fig.~\ref{f:map}(b) shows the residual obtained by
subtracting two independent surface deviation measurements.
The rms of this residual is $\sim70~\mu$m. Therefore, the measurement accuracy
is estimated to be $\sim50~\mu$m.

%%%%%%%%%%%%%%%%%%%%%%%%%%%%%%%%%%%%%%%%%%%%%%%%%%%%%%%%%
\section*{Acknowledgment}
\small We would like to thank the Radio Astronomy Lab (RAL), RRI, for providing
the components used in the receiver chain. We would also like to thank Prof
V.~Lakshminarayanan (LC Lab), for sharing the SRS Lockin Amplifier.

%%%%%%%%%%%%%%%%% Figures %%%%%%%%%%%%%%%%%

%\begin{figure}[]
%\centering
%\epsfig{figure=Al_Panel_worked_diff.eps,width=\linewidth}
%\caption{\small Difference map showing measurement accuracy. \normalsize }
%\label{f:diff_map}
%\end{figure}

%\begin{figure}[]
%\centering
%\epsfig{figure=HoloMap.eps,width=0.48\linewidth}
%\caption{\small Scheme for raster scan. \normalsize }
%\label{f:rasterscan}
%\end{figure}

%%%%%%%%%%%%%%%%%%%%%%%%%%%%%%%%%%%%%%%%%%%%%%%%%%%%%%%%%
%\vspace{+20mm}

\end{document}